# A nanometer-scale optical electrometer

A. N. Vamivakas[1], Y. Zhao[1,2], S. Fält[3], A. Badolato[4], J. M. Taylor[5,6] & M. Atatüre[1]

[1]*Cavendish Laboratory, University of Cambridge, JJ Thomson Avenue, Cambridge CB3 0HE, UK*

[2]*Physikalisches Institut, Ruprecht-Karls-Universität Heidelberg, Philosophenweg 12, 69120 Heidelberg, Germany*

[3]*Sol Voltaics AB, Scheelevägen 17, Ideon Science Park, 223 70 Lund, Sweden*

[4] *Department of Physics and Astronomy, University of Rochester, Rochester, New York 14627, USA*

[5]*Department of Physics, Massachusetts Institute of Technology, 77 Massachusetts Ave, Cambridge, MA 02139, USA*

[6]*Joint Quantum Institute/National Institute of Standards of Technology, 100 Bureau Drive MS 8423, Gaithersburg, MD 20899*

**Self-assembled semiconductor quantum dots show remarkable optical and spin coherence properties, which have lead to a concerted research effort examining their potential as a quantum bit for quantum information science[1-6]. Here, we present an alternative application for such devices, exploiting recent achievements of charge occupation control and the spectral tunability of the optical emission of quantum dots by electric fields[7] to demonstrate high-sensitivity electric field measurement. In contrast to existing nanometer-scale electric field sensors, such as single electron transistors[8-11] and mechanical resonators[12,13], our approach relies on homodyning light resonantly Rayleigh scattered from a**

**quantum dot transition with the excitation laser and phase sensitive lock-in detection. This offers both static and transient field detection ability with high bandwidth operation and near unity quantum efficiency. Our theoretical estimation of the static field sensitivity for typical parameters, 0.5 V/m/√Hz, compares favorably to the theoretical limit for single electron transistor-based electrometers. The sensitivity level of 5 V/m/√Hz we report in this work, which corresponds to 6.4 * $10^{-6}$ e/√Hz at a distance of 12 nm, is worse than this theoretical estimate, yet higher than any other result attained at 4.2 K or higher operation temperature.**

The electric field dependence of self-assembled semiconductor quantum dots (QDs) originates from an inherent displacement of the electron and hole wave functions within the confinement potential of the QD volume. The small displacement, typically around 0.5 nm, is sufficient to generate a non-zero expectation value for the exciton dipole moment[14]. This permanent dipole moment leads to a linear Stark shift ~ $\alpha \boldsymbol{n} \cdot \boldsymbol{E}$ of the transition under an applied electric field $\boldsymbol{E}$, where $\alpha$ ~ 0.028 MHz (V/m)$^{-1}$ is the permanent dipole moment in a typical device and $\boldsymbol{n}$ is the QD growth direction. This large dipole moment results in the QD exciton transition being sensitive to the presence of local electric charge. For example, the presence of a single electron 12 nm from a QD results in an excitonic spectral shift that is 2-orders of magnitude larger than the transition's linewidth of ~ 250 MHz[15]. This corresponding spectral shift can be detected using two simultaneous laser homodyne signals arising from the interference of the excitation laser with the QD transition forward and backward scattered light, i.e. differential transmission (DT) and reflection (DR)[16,17]. Since the local electric field is measured via light scattering, the bandwidth of our electrometer is fundamentally limited by the rate at which the exciton transition scatters photons – the spontaneous emission rate $\gamma_{sp}$.

To demonstrate the suitability of DT and DR signals for static and transient electric field sensing we carried out a proof-of-principle experiment with a single self-assembled Indium Arsenide QD in Gallium Arsenide embedded in a Schottky diode structure [see Methods]. For all experiments the voltage across the diode enables external control of the excitonic transition energy via the Stark effect and the diode is operated at voltage such that no current passes through the QD. Figure 1a illustrates the experimental setup used for this purpose. The excitation laser generates Rayleigh scattered photons when tuned to a QD transition. Two detectors access the forward and backward scattered fields enabling a simultaneous measurement of the DT and DR signals. Lock-in detection based on gate voltage modulation suppresses low frequency detector and laser noise. Figure 1b presents DT from a singly charged QD as a function of voltage and excitation laser frequency revealing a linear Stark shift response of 140 MHz/mV. The DT signal reproduces a Lorentzian lineshape if the voltage modulation amplitude results in larger than QD linewidth spectral shifts. To detect small changes in the QD's resonance frequency—necessary for sensing electric fields— a small (sub-linewidth) voltage amplitude modulation, which gives the derivative of the lineshape, is applied. Figure 1c illustrates how DT can be used in conjunction with lock-in detection to sense static electric fields. The experimentally measured DT signal for sub-linewidth voltage modulation is displayed in Fig 1d. Here the voltage modulation amplitude is 1 mV corresponding to 0.14 GHz on the abscissa. We emphasize this is an AC measurement, however the small voltage modulation enables DC and quasi-static field detection.

Figure 1e illustrates how the DR signal can be used in conjunction with lock-in detection to measure transient electric fields. The DR lineshape is anti-symmetric due to the relative phase of the reflected laser and the resonant Rayleigh scattering and becomes similar to the second derivative of the original symmetric Lorentzian when the voltage modulation amplitude is reduced to 1mV. In this case, the DR signal has no response to a static electric field, but is sensitive to transient electric fields with a frequency component near the modulation frequency. Figure 1f displays the measured DR signal in this regime. The deviation of the lineshape from the second derivative of an ideal Lorentzian originates from the phase between the laser and the Rayleigh scattered light, as determined by the distance between the QD and the top surface of the sample. Nevertheless, this lineshape still provides the response we seek to a transient electric field. Simultaneous measurement of DR and DT thus allows for both AC and DC detection.

Our detection approach is limited fundamentally by the optical theorem which relates the maximum amount of scattered and fluorescent light to the transition's radiative lifetime. Further, it achieves unit quantum efficiency since every photon scattered from the QD leads to DT/DR signal. Specifically, the rate of photons incident on the QD is given by the rate of photons incident on the detector in the absence of the QD scaled by $p = 4/(3nNA)^2$, with a Rabi frequency

$$\Omega = \sqrt{\frac{P}{\hbar\omega}\frac{\gamma_{sp}}{p}} \qquad (1)$$

where $n$ is the refractive index, $NA$ the lensing system numerical aperture $P$ the laser power, $\omega$ the laser frequency and $\gamma_{sp}$ the spontaneous emission rate. Assuming a Lorentzian lineshape for the underlying transition with a linewidth $\Gamma$, the rate of photons incident on the detector is

$$\dot{n} = \frac{\Omega^2}{\gamma_{sp}}\left(p - \frac{2\gamma_{sp}\Gamma}{\Delta^2 + \Gamma^2 + 4\Omega^2\Gamma/\gamma_{sp}}\right) \qquad (2)$$

where $\Delta$ is the electric field dependent spectral detuning of the laser from the QD transition. The corresponding shot-noise is $\sqrt{\dot{n}t}$ so the signal-to-noise improves by the square root of averaging time leading to our electric field sensitivity of

$$\eta = \varepsilon^{-1/2}\alpha^{-1}\frac{\sqrt{\dot{n}}}{\partial\dot{n}/\partial\Delta} \qquad (3)$$

This is maximal for $\Delta = \Gamma/\sqrt{2}$ and $\Omega = \sqrt{\gamma_{sp}\Gamma/8}$ giving

$$\eta_{max} = \varepsilon^{-1/2}\alpha^{-1}\frac{4\Gamma^{3/2}\sqrt{p}}{\gamma_{sp}}. \qquad (4)$$

For our experimental parameters the maximum sensitivity is ~ 0.5 V/m/√Hz.

    To characterize the QD as a DC electrometer at 0 mV gate-voltage shift in Fig. 1b the square wave modulation peak-to-peak voltage amplitude is fixed to 1 mV. The DT signal is recorded for 8 s with 1 ms resolution for a range of laser frequencies. A trace of the mean DT signal is presented in Fig. 1d when the transition is driven slightly above saturation. Evident in Fig. 1d is the strong variation in the mean DT signal around resonance – the DC electrometer operating point – as compared to frequency shifts greater than ±0.25 GHz. Figure 2a displays the standard deviation of time traces from 3 spectral locations as the measurement time constant is increased. Measurements performed away from the QD resonance (0.75 GHz laser detuning) in Fig. 2a display that the resulting noise (open blue circles) reduces to the square root of the averaging time (black line in the inset) as expected from white noise within our measurement bandwidth. The same behavior is observed when the laser is detuned 0.25 GHz from resonance (red squares), where, as can be seen in Fig. 1d, the static electric field dependence is expected to be suppressed while still detecting DT signal. These measurements mark the experimentally obtainable levels of noise floor in our detection, which set a lower bound on the electric field sensitivity we can expect from our sensor.

However, on resonance, the noise behavior deviates from the square-root dependence on measurement time indicating a departure from purely white noise. This behavior arises from charge dynamics in the QD environment and it is difficult to distinguish these fluctuations from an external electric field we wish to detect. We need to consider this noise floor in order to identify the detection threshold of our current device for a given measurement time. An alternative perspective is that our QD as a field sensor is already performing an electric field measurement of its environment and this is the first direct measurement of the noise behavior for these low frequency charge fluctuations in the vicinity of self-assembled QDs and will be further studied elsewhere.

To quantify the DC electrometer sensitivity we use the data presented in Fig. 2. The top panel of Fig. 2b presents the mean DT signal recorded for each laser frequency (symbols correspond to the noise measurements presented in Fig. 2a). Panel II in Fig. 2b is the mean DT signal derivative identifying the region of high gradient useful for sensing. To determine the sensitivity spectral variation (in the absence of charge fluctuations) we use the noise measurements from 0.75 GHz detuning in Eq. 2 and present $\eta$ in Panel III. A sensitivity of 5 V/m/√Hz is obtained across the 100 MHz window around the resonance. Alternatively, for 80 ms measurement time this leads to a detection threshold of 110 V/m when the charge fluctuations are considered inherent to the device. Figure 2c presents the laser power dependence of $\eta$. We find the maximum static field detection threshold for 80 ms measurement time is $110 \pm 20$ V/m at 1.5 nW which corresponds to the saturation ($\Omega \approx \Gamma$).

Figure 3a presents fixed gate voltage DR spectroscopy of the QD transition for a range of voltage modulation amplitudes ($V_{pk-pk}$). We identify the operating modulation peak-to-peak amplitude by analyzing the DR line cut (inset) in Fig. 3a (solid black line). The arrow indicates the 2-mV operating amplitude, which provides the gradient sensitive to an oscillatory field at the lock-in frequency. Similar to the analysis presented in Fig.

2b, the noise follows the square root dependence on measurement time when the voltage modulation is outside the electric field sensitive high-gradient regime (blue open triangles). The noise behavior also exhibits a deviation in the field sensitive region (open red triangles) similar to the DC electrometer (open orange diamonds). This is a direct outcome of suppressed (by a factor of 3) but not fully rejected low frequency charge dynamics discussed above. One advantage of our ability to simultaneously record the DR and DT signals is that residual coupling of low frequency signals can be subtracted from the DR signal allowing a significantly higher level of noise rejection for the AC electrometer.

Determination of the AC electrometer's sensitivity $\eta$ is accomplished by applying a 1.908 KHz sinusoidal field to the gate for a range of amplitudes. The laser is on resonance at saturation power and 8s of data is acquired with 0.5 μs temporal resolution. Panel I of Fig. 3c plots the mean DR signal as the modulation amplitude of the additional field is varied. The mean DR signal and its derivative (Panel II) are used to calculate the sensitivity in Panel III in accordance with Eq. 3 using the noise level obtained in the absence of residual DC coupling. The measured sensitivity $\eta$, 140 V/m/√Hz, depends on the modulation amplitude of the sinusoidal voltage and it reduces to 14 V/m/√Hz for an input sinusoidal field amplitude of 1.5 mV due to the nonlinear gradient of this transition response (Panel I of Fig. 3c). A wavelength-optimized distance between the QD and the top surface should provide a more linear response whereby the sensitivity would be input-amplitude independent. For our system we find 140 V/m/√Hz sensitivity for an electric field oscillating at 1.908 kHz with a resolution of 10 Hz determined solely by the lock-in electronics. If we only use the DR signal without rejection of any residual low frequency coupling, we see that the detection threshold for 80 ms measurement time is $10^3$ V/m at vanishing input signal amplitude and this value marks the figure of merit for the whole device as an AC electrometer.

So far we have tested the sensitivity of our device using macroscopic, externally applied electric fields. However, for a single quantum dot, the detection ability does not depend on the overall electric field through a volume, as it might in a capacitive detector, but rather it only depends on the field at the location of the QD confined to 20 nm such that the dot maintains its optical properties. Thus, our device is best applied for situations with high electric field densities but low total electrical energy, such as single charge detection or defect detection in nearby materials and surfaces. In this context, the best available electrometers at the moment are single electron transistors.

The results presented show that the linear Stark shift of the QD emission spectrum can be exploited to determine the presence of static and transient electric fields simultaneously with high sensitivity on nanometer length scales. This corresponds to $6.4 * 10^{-6}$ e/√Hz sensitivity for monitoring electron charging statistics of a neighboring QD at a distance of 12 nm[15]. Conditional on reasonable sample structure technical improvements, light extraction efficiency[18], and comparator-based electronics to further suppress residual coupling between DT and DR, our system has the potential to reach the current state-of-the-art single-electron-transistor electrometer at tens of Kelvins without needing the circuitry required for 1-D and SET detectors. QDs with stronger Stark shift coefficients will also improve the device performance. With currently available detection electronics, our electrometer design can cover an operation range from DC to 100s of MHz limited ultimately by the optical decay rate $\gamma_{sp}$. Another potential advantage of this system is that many QDs spaced closely with different optical frequencies around a complex system can be probed independently enabling a high-resolution two dimensional map of the electric field. Further, the back action of our device on the electron being measured originates from the dipolar field due to the strong confinement of the charge neutral exciton, therefore back-action in our system is considerably smaller than others. Finally, we note that our system can also be operated as an optical magnetometer[19,20,21,22] to sense magnetic fields via the linear Zeeman shift of 30 GHz/T for the QD transition, yielding a sensitivity of ~$5 * 10^{-6}$ T/√Hz.

**Methods**

The InAs/GaAs quantum dots studied were grown by molecular beam epitaxy and embedded in a Schottky diode heterostructure. The device structure is described in Ref. 23. For the purpose of this study only single QDs were investigated. An applied gate bias allows deterministic loading of individual electrons from $n^+$ layer. The sample is housed in a magneto-optical bath cryostat and cooled to 4.2 K. A cubic zirconia solid immersion lens is utilized to improve both the light focusing and light gathering power of the fiber-based confocal microscope. The differential transmission and reflection spectroscopy technique use a scanable single mode diode laser with 1.2-MHz frequency and 0.5% power stabilization.

**Acknowledgements** This work was supported by grants and funds from the University of Cambridge, EPSRC Science and Innovation Awards, EPSRC grant number EP/ G000883/1 and the Cambridge-MIT exchange grant P019. Y.Z. is supported by the A. v. Humboldt Foundation and LGFG. The authors thank Charles Smith for useful discussions.

**Figure Captions**

**Fig. 1 Experimental set-up, QD $X^{1-}$ spectroscopy and operation of QD $X^{1-}$ as an AC- and DC-electrometer. a,** Illustration of the experimental apparatus used for optical AC- and DC-electrometry. The quantum dot (QD) sample is mounted on a Si photodiode and both sit on x-y-z positioners in a liquid helium bath cryostat. Through 1 arm of the confocal microscope a single mode laser is focused onto the sample. Laser light and Rayleigh scattered light reaching the Si photodiode underneath the sample is used for DC electrometry whereas laser light reflected from the sample surface and Rayleigh scattered light collected in the second arm of the confocal microscope is used for AC electrometry. **b,** Differential transmission (DT) spectroscopy of $X^{1-}$. For each DC gate voltage offset the laser frequency is scanned. The laser power equals the saturation power and the QD gate voltage is modulated by a square wave with peak-to-peak amplitude of 100 mV to enable phase sensitive detection of the DT signal. **c,** Illustration of how a Lorentzian response can be exploited with phase sensitive detection for DC electrometry. The top panel illustrates 1 cycle of the lock-in detector when the laser, vertical arrow, is tuned to the point of highest slope for the Lorentzian response reporting a signal level s. The lock-in output in this case is 0. The lower panel is 1 cycle of the lock-in detector, but here the laser is detuned from the highest slope point due to the presence of an applied electric field. The lock-in reports a non-zero voltage that reveals the magnitude of the applied electric field. Note the transition symmetry and lock-in phase makes the measured voltage insensitive to transient electric fields. **d,** DT laser frequency scan acquired with a square wave modulation of peak-to-peak amplitude equal to 1 mV. The measured DT signal is the derivative of the Lorentzian response. **e,** Illustration of how a dispersive response can be exploited with phase sensitive detection for AC electrometry. The top panel illustrates 1 cycle of the lock-in detector when the laser, vertical arrow, is tuned to a point such that the lock-in reports a signal level -2s. The lower panel is 1 cycle of the lock-in detector, but here the

laser probes multiple signal values of the transition due to the presence of an applied transient electric field at the lock-in frequency. The lock-in reports an additional voltage that reveals the magnitude of the applied electric field. Note the transition symmetry and lock-in phase makes the measured voltage insensitive to static electric fields. **f,** DR laser frequency scan acquired with a square wave modulation of peak-to-peak amplitude equal to 2 mV. The measured DR signal is the derivative of the dispersive response.

**Fig. 2 DC electrometry. a,** A log-log plot of the variation of measurement noise, quantified by the standard deviation, with measurement time constant. Each data point is determined by subdividing the measured time trace into a collection of equal size time bins, finding the mean of each time bin and then finding the standard deviation of the collection of means. The open orange diamonds correspond to 0 frequency shift in panel I of Fig. 2b, the open red squares 0.25 Frequency shift and the open blue circles 0.75 frequency shift. At the DC electrometer operating point – 0 frequency shift – the noise magnitude is largest. Inset: Same noise data, but all curves are normalized to 1 at 10 ms time constant. The black line plots the inverse square-root of the measurement time constant. For spectral locations away from the DC electrometer operating point (0 Frequency shift in panel I of Fig. 2b) the noise exhibits the expected time constant dependence. When the laser is at the electrometer operating spectral region the measured noise is colored as indicated by the departure of the data from the black line. **b,** Panel I: Differential transmission signal when the laser power is slightly above saturation and the peak-to-peak amplitude of the square wave modulation is 1 mV. Each point is determined from the mean of 8s worth of data acquired with 1 ms time bins. Panel II: Numerical derivative of panel I. Panel III: Laser frequency dependence of the DC sensitivity. To calculate the sensitivity the noise for the 0.75 GHz time trace

is used. **c,** The laser power dependence of the measured sensitivity when the laser is resonant with the $X^{1-}$ transition. We find the best sensitivity is 5 V/m/√Hz.

**Fig. 3 AC electrometry. a,** Differential reflection (DR) spectroscopy of the $X^{1-}$ as the square wave modulation peak-to-peak voltage is varied. For each peak-to-peak voltage the laser frequency is scanned across the transition resonance. Inset: DR signal along the black dashed line. The vertical arrow identifies the square wave modulation peak-to-peak voltage of 2 mV used for the data presented in panels **c** and **d. b,** A log-log plot of the variation of measurement noise, quantified by the standard deviation, with measurement time constant. Each data point is determined by subdividing the measured time trace into a collection of equal size time bins, finding the mean of each time bin and then finding the standard deviation of the collection of means. The open orange diamonds correspond to the DC electrometer operating point presented in Fig. 2a. The up red (down blue) triangles correspond to AC noise when the AC electrometer is characterized at (away) from its operating point. Inset: Same noise data, but all curves are normalized to 1 at 10 ms time constant. The black line plots the inverse square-root of the measurement time constant. For configurations away from the AC electrometer operating point the noise exhibits the expected time constant dependence. When the system is at the AC or DC electrometer operating point the measured noise is colored as indicated by the departure of the data from the black line. **c,** Panel I: DR signal when the laser power is slightly above saturation, the peak-to-peak amplitude of the square wave modulation is 2 mV and the amplitude of an additional sine wave applied to the QD gate is varied. Each point is determined from the mean of 8s worth of data acquired with 0.5 μs time bins. The error bars correspond to 1 standard deviation for one second of averaging time. Panel II: Numerical derivative of panel I. Panel III: Sine wave peak-to-peak voltage amplitude dependence of the AC sensitivity. To calculate the sensitivity the standard deviation of each time trace (for each sine wave peak-to-peak voltage amplitude) is evaluated by subdividing the full time trace into 80 ms bins, finding the average of each bin and then evaluating the standard deviation of this collection of means. The horizontal black line identifies the AC sensitivity when the sinusoid's peak-to-peak voltage amplitude approaches 0.

FIGURE 1

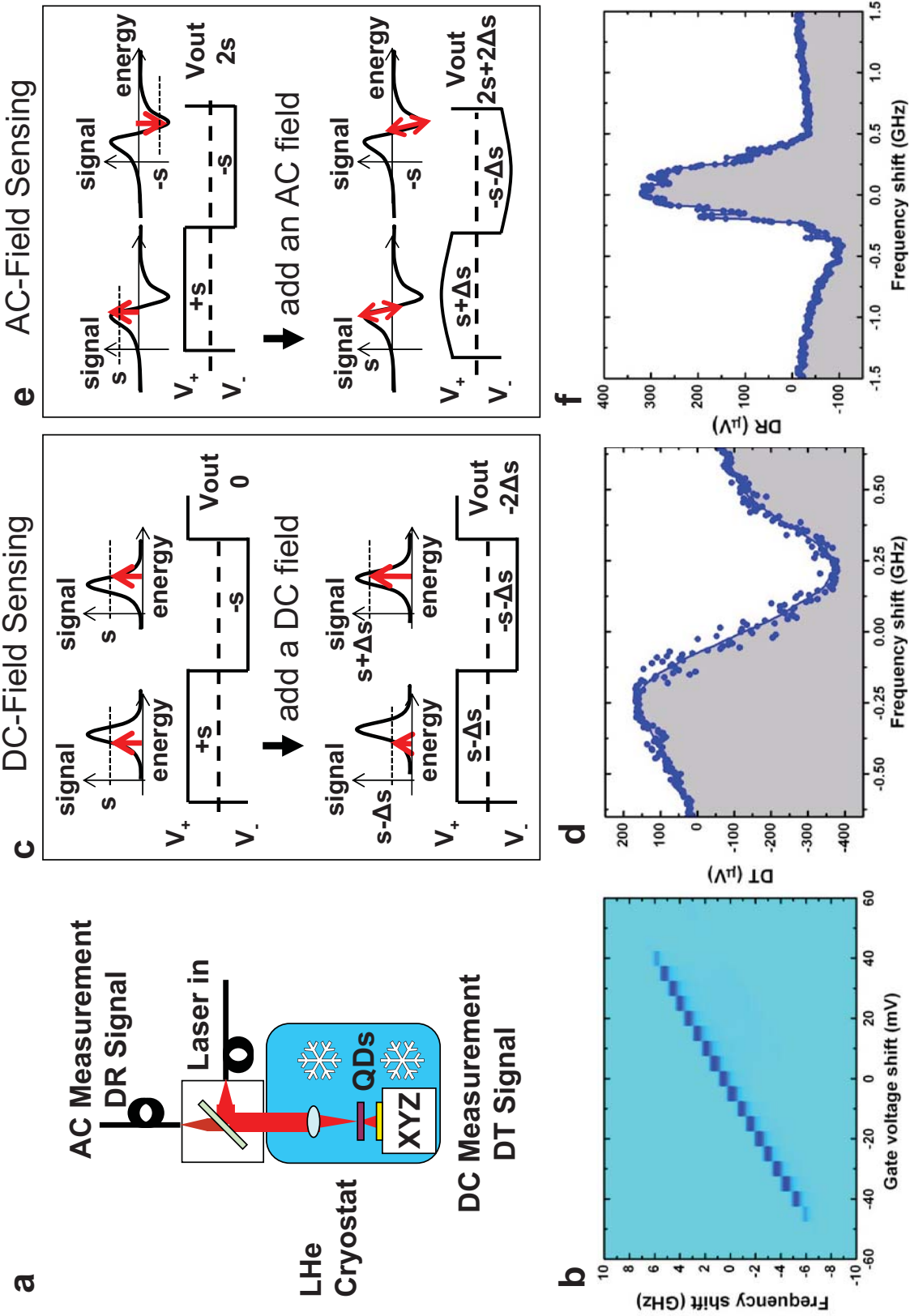

FIGURE 2

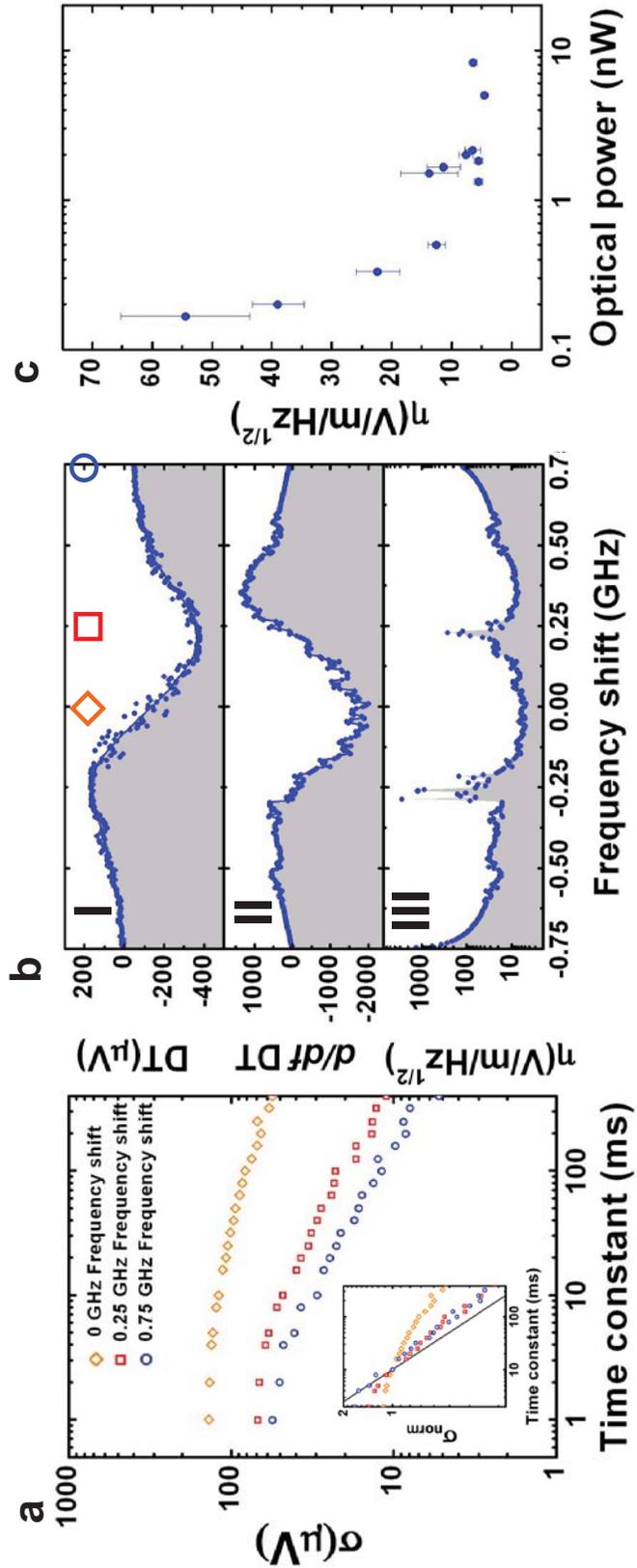

FIGURE 3

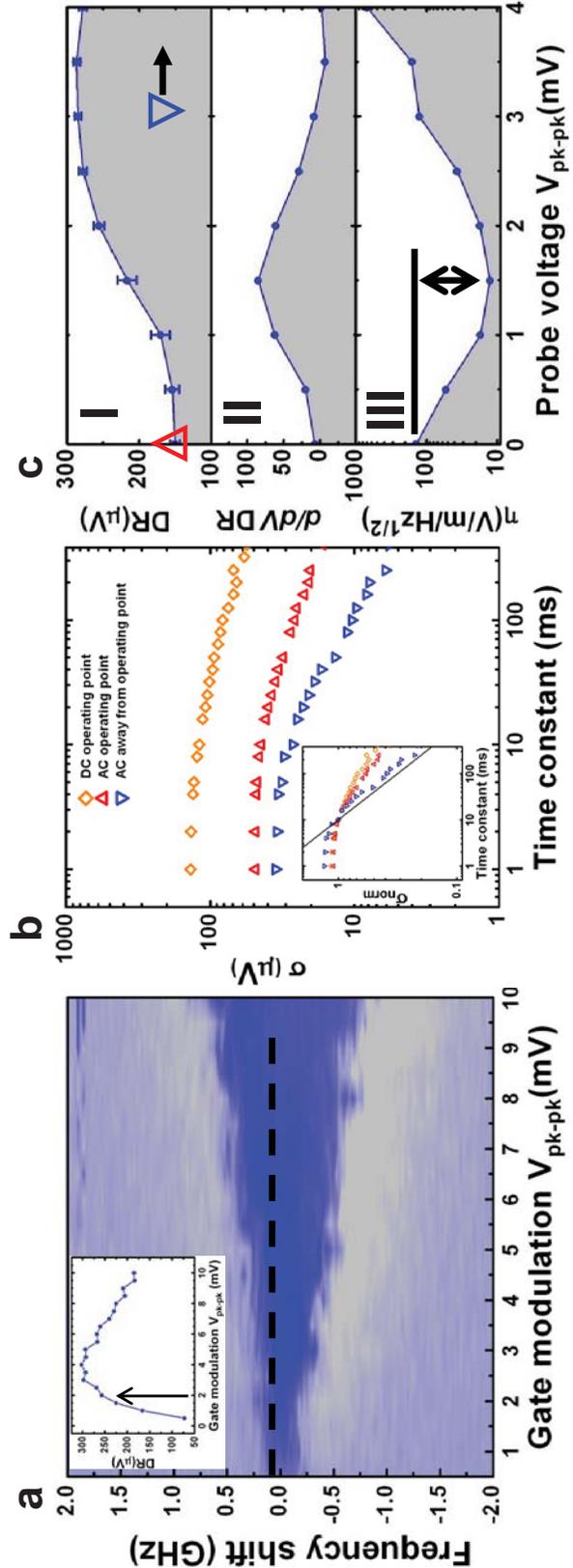